\documentclass[longauth]{aa} 
\usepackage{graphicx}
\usepackage{txfonts}

\usepackage{amsmath}
\usepackage{amsmath}
\usepackage{savesym}
\savesymbol{iint}
\usepackage{wasysym}
\restoresymbol{WAS}{iint}
\usepackage{natbib}
\bibpunct{(}{)}{;}{a}{}{,} 
\begin{document}
\newcommand{\smst}{{\it sample ${\cal M}$ST}} 
\newcommand{\smcor}{{\it sample ${\cal M}$COR}}
\newcommand{\st}{{\it ${\cal M}$ST}} 
\newcommand{\mcor}{{\it ${\cal M}$COR}} 
\newcommand{\masq}{magn \arcsec$^{-2}$}
\newcommand{\Jykms}{Jy!\,km\,s$^{-1}$}
\newcommand{\mjyb}{mJy beam$^{-1}$}
\newcommand{\HO}{$H_0$}
\newcommand{\kms}{km\,s$^{-1}$}
\newcommand{\acm}{cm$^{-2}$}
\newcommand{\teff}{$T_{eff}$} 
\newcommand{\kmsmpc}{km\,s$^{-1}$\,Mpc$^{-1}$}
\newcommand{\mjb}{mJy beam$^{-1}$}
\newcommand{\jb}{Jy beam$^{-1}$}
\newcommand{\mjbc}{mJy beam$^{-1}$ channel$^{-1}$}
\newcommand{\jykms}{Jy km s$^{-1}$}
\newcommand{\logm}{log{\small ($\cal M_* / M_\odot$)}} 
\newcommand{\msun}{$\cal M_\odot$}
\newcommand{\msunyr}{$\cal M_\odot$\,yr$^{-1}$}
\newcommand{\lsun}{{$L_{\odot,B$}}}
\newcommand{\lb}{{L$_{\rm B}$}}
\newcommand{\mlsun}{{\rm M}$_\odot/{\rm L}_{\rm B_{\odot}$}}
\newcommand{\kmsMpc}{km s$^{-1}$ Mpc$^{-1}$}
\newcommand{\hi}{H{\small I}}
\newcommand{\MHI}{{$\cal M}_{\rm HI}$}
\newcommand{\m}{\hbox{$^{\rm m}$}}
\newcommand{\s}{\hbox{$^{\rm s}$}}
\newcommand{\h}{\hbox{$^{\rm h}$}}
\newcommand{\dn}{D$_{\rm n}$4000}
\newcommand{\halpha}{H$\alpha$}
\newcommand{\ewhalpha}{EW(H$\alpha$)}
\newcommand{\hbeta}{H$\beta$}
\newcommand{\ewhbeta}{EW(H$\beta$)}
\newcommand{\hgamma}{H$\gamma$}
\newcommand{\hepsilon}{H$\epsilon$}
\newcommand{\hdelta}{H$\delta$}
\newcommand{\hdeltaA}{H$\delta_{A}$}
\newcommand{\ewhdelta}{EW(H$\delta$)}
\newcommand{\hds}{H$\delta$~strong}
\newcommand{\oii}{[O{\small II}]}
\newcommand{\ewoii}{EW[O{\small II}]}
\newcommand{\oiii}{[O{\small III}]}
\newcommand{\oiiia}{[O{\small III}]${\lambda}$5007}
\newcommand{\oiiib}{[O{\small III}]${\lambda}$4959}
\newcommand{\ewoiii}{EW[O{\small III}]}
\newcommand{\neiii}{[Ne{\small III}]} 
\newcommand{\caii}{Ca{\small II}} 
\newcommand{\ie}{{i.e.\,}}
\newcommand{\eg}{{e.g.\,}}
\newcommand{\psb}{{post-starburst}}
\newcommand{\Psb}{{Post-starburst}}
\newcommand{\ka}{{$k+a$}}
\newcommand{\Ka}{{$K+a$}}
\newcommand{\chandra}{{\it Chandra\/}}
\newcommand{\xmm}{{XMM-{\it Newton\/}}}
\newcommand{\cgs}{{${\rm erg~cm}^{-2}~{\rm s}^{-1}$}}
\newcommand{\lum}{{\rm erg~s$^{-1}$}}
\newcommand{\rk}{(r-K)}
\newcommand{\nev}{[Ne{\small V}]} 
\newcommand{\nevl}{[Ne{\small V}]${\lambda}$3426} 
\newcommand{\hii}{[H{\small II}]} 
\newcommand{\micron}{$\mu$m}
\newcommand{\nh}{${\rm N_\mathrm{H}}$}
\newcommand{\ergs}{\ifmmode {\rm\,ergs\,s^{-1}}\else    ${\rm\,ergs\,s^{-1}}$\fi}

  \title{The VIMOS Public Extragalactic Redshift Survey (VIPERS). \\ AGN feedback in [NeV] emitters \thanks{based on observations collected at the European Southern Observatory, Cerro Paranal, Chile, using the Very Large Telescope under programs 182.A-0886 and partly 070.A-9007. Also based on observations obtained with MegaPrime/MegaCam, a joint project of CFHT and CEA/DAPNIA, at the Canada-France-Hawaii Telescope (CFHT), which is operated by the National Research Council (NRC) of Canada, the Institut National des Sciences de l'Univers of the Centre National de la Recherche Scientifique (CNRS) of France, and the University of Hawaii. This work is based in part on data products produced at TERAPIX and the Canadian Astronomy Data Centre as part of the Canada-France-Hawaii Telescope Legacy Survey, a collaborative project of NRC and CNRS. The VIPERS web site is http://www.vipers.inaf.it/.}}
  
  \titlerunning{AGN feedback on star formation}

\author{
D.~Vergani\inst{\ref{iasf-bo}}
\and B.~Garilli\inst{\ref{iasf-mi}}          
\and M.~Polletta\inst{\ref{iasf-mi},\ref{marseille-uni},\ref{toulouse}}
\and P.~Franzetti\inst{\ref{iasf-mi}}   
\and M.~Scodeggio\inst{\ref{iasf-mi}}
\and G.~Zamorani\inst{\ref{oabo}}
\and C.~P.~Haines\inst{\ref{brera}}
\and M.~Bolzonella\inst{\ref{oabo}} 
\and L.~Guzzo\inst{\ref{brera},\ref{unimi}}     
\and B.~R.~Granett\inst{\ref{brera},\ref{unimi}}                                                      
\and S.~de la Torre\inst{\ref{lam}}       
\and U.~Abbas\inst{\ref{oa-to}}
\and C.~Adami\inst{\ref{lam}}
\and D.~Bottini\inst{\ref{iasf-mi}}
\and A.~Cappi\inst{\ref{oabo},\ref{nice}}
\and O.~Cucciati\inst{\ref{oabo},\ref{unibo}}           
\and I.~Davidzon\inst{\ref{lam},\ref{oabo}}   
\and G.~De Lucia\inst{\ref{oats}}
\and A.~Fritz\inst{\ref{iasf-mi}}      
\and A. Gargiulo\inst{\ref{iasf-mi}}
\and A.~J.~Hawken\inst{\ref{brera},\ref{unimi}} 
\and A.~Iovino\inst{\ref{brera}}
\and J.~Krywult\inst{\ref{kielce}}
\and V.~Le Brun\inst{\ref{lam}}
\and O.~Le F\`evre\inst{\ref{lam}}
\and D.~Maccagni\inst{\ref{iasf-mi}}
\and K.~Ma{\l}ek\inst{\ref{warsaw-nucl}}
\and F.~Marulli\inst{\ref{unibo},\ref{infn-bo},\ref{oabo}} 
\and A.~Pollo\inst{\ref{warsaw-nucl},\ref{krakow}}
\and L.A.M.~Tasca\inst{\ref{lam}}
\and R.~Tojeiro\inst{\ref{st-andrews}}  
\and A.~Zanichelli\inst{\ref{ira-bo}}
\and S.~Arnouts\inst{\ref{lam}} 
\and J.~Bel\inst{\ref{cpt}}
\and E.~Branchini\inst{\ref{roma3},\ref{infn-roma3},\ref{oa-roma}}
\and J.~Coupon\inst{\ref{geneva}}
\and O.~Ilbert\inst{\ref{lam}}
\and T.~Moutard\inst{\ref{halifax},\ref{lam}}  
\and L.~Moscardini\inst{\ref{unibo},\ref{infn-bo},\ref{oabo}}
} 
\institute{
INAF - Istituto di Astrofisica Spaziale e Fisica Cosmica Bologna, via P. Gobetti 101, I-40129 Bologna, Italy, \email{vergani@iasfbo.inaf.it}  \label{iasf-bo}
\and INAF - Istituto di Astrofisica Spaziale e Fisica Cosmica Milano, via E. Bassini 15, 20133 Milano, Italy \label{iasf-mi}
\and Aix-Marseille Universit\`{e}, Jardin du Pharo, 58 bd Charles Livon, F-13284 Marseille cedex 7, France \label{marseille-uni}
\and IRAP,  9 av. du colonel Roche, BP 44346, F-31028 Toulouse cedex 4, France \label{toulouse} 
\and INAF - Osservatorio Astronomico di Bologna, via P. Gobetti 93/3, I-40129, Bologna, Italy \label{oabo}
\and INAF - Osservatorio Astronomico di Brera, via Brera 28, 20122 Milano --  via E. Bianchi 46, 23807 Merate, Italy \label{brera}
\and  Universit\`{a} degli Studi di Milano, via G. Celoria 16, 20133 Milano, Italy \label{unimi}
\and Aix Marseille Univ, CNRS, LAM, Laboratoire d'Astrophysique de Marseille, Marseille, France  \label{lam}
\and INAF - Osservatorio Astrofisico di Torino, 10025 Pino Torinese, Italy \label{oa-to}
\and Laboratoire Lagrange, UMR7293, Universit\'e de Nice Sophia Antipolis, CNRS, Observatoire de la C\^ote d'Azur, 06300 Nice, France \label{nice}
\and Dipartimento di Fisica e Astronomia - Alma Mater Studiorum Universit\`{a} di Bologna, via P. Gobetti 93/2, I-40129 Bologna, Italy \label{unibo}
\and INAF - Osservatorio Astronomico di Trieste, via G. B. Tiepolo 11, 34143 Trieste, Italy \label{oats}
\and Institute of Physics, Jan Kochanowski University, ul. Swietokrzyska 15, 25-406 Kielce, Poland \label{kielce}
\and National Centre for Nuclear Research, ul. Hoza 69, 00-681 Warszawa, Poland \label{warsaw-nucl}
\and INFN, Sezione di Bologna, viale Berti Pichat 6/2, I-40127 Bologna, Italy \label{infn-bo}
\and Astronomical Observatory of the Jagiellonian University, Orla 171, 30-001 Cracow, Poland \label{krakow}
\and School of Physics and Astronomy, University of St Andrews, St Andrews KY16 9SS, UK \label{st-andrews}
\and INAF - Istituto di Radioastronomia, via P. Gobetti 101, I-40129, Bologna, Italy \label{ira-bo}
\and Aix Marseille Univ, Univ Toulon, CNRS, CPT, Marseille, France \label{cpt}
\and Dipartimento di Matematica e Fisica, Universit\`{a} degli Studi Roma Tre, via della Vasca Navale 84, 00146 Roma, Italy\label{roma3} 
\and INFN, Sezione di Roma Tre, via della Vasca Navale 84, I-00146 Roma, Italy \label{infn-roma3}
\and INAF - Osservatorio Astronomico di Roma, via Frascati 33, I-00040 Monte Porzio Catone (RM), Italy \label{oa-roma}
\and Department of Astronomy, University of Geneva, ch. d'Ecogia 16, 1290 Versoix, Switzerland \label{geneva}
\and Department of Astronomy $\&$ Physics, Saint Mary's University, 923 Robie Street, Halifax, Nova Scotia, B3H 3C3, Canada \label{halifax}
}

  \offprints{\mbox{D. Vergani}, \email{vergani@iasfbo.inaf.it}}

 \date{Received ---; accepted ---}
 
 \abstract {Using an unconventional single line diagnostic that unambiguously identifies AGNs in composite galaxies we report statistical differences in the properties (stellar age, \oii\ luminosity, colour) between active and inactive galaxies at $0.62 < z < 1.2 $ extracted from the VIMOS Public Extragalactic Redshift Survey (VIPERS). The nuclear activity is probed by the high-ionization \nevl\ emission line and along with their parent samples, the active galaxies are properly selected according to their stellar mass, redshift, and NUVrK colour distributions. We report younger underlying stellar ages and higher \oii\ luminosities of active galaxies in the green valley and in the blue cloud compared to control samples. We observe higher fractions of green galaxies hosting AGN activity at progressively bluer \rk\ colours. Depending on the location of the host galaxy in the NUVrK colour diagram we find higher AGN fractions in massive blue galaxies and in the least massive red galaxies, in agreement with the picture that black holes vary their properties when hosted in either star-forming or passive galaxies.  Exactly where the fast quenching processes are expected to play a role, we identify a novel class of active galaxies in the blue cloud with signatures typical for a suddenly suppression of their star formation activity after a burst happening in the recent past (less than 200-300\,Myr earlier). Their optical spectra resemble those of post-starburst galaxies, that would never be identified in a spectroscopic search using classical post-starburst selection techniques. Broadly, these active galaxies selected on the \nevl\ line are not commonly represented in shallow X-ray, mid-IR, or classical line diagnostics. If we consider that our results are limited by the shallow observational limits and rapid AGN variability, the impact of AGN feedback on galaxy formation and evolution may represent an important channel of fast-transiting galaxies moving to the red sequence.}

\maketitle

\section{Introduction}
\label{sec:intro}

Various mechanisms have been proposed to explain the ceasing of star
formation activity in galaxies. They are cluster-related mechanisms,
such as ram-pressure gas stripping, harassment, or strangulation
\citep{gunn72, larson80, balogh00}, including events like galaxy
mergers and interactions \citep{toomre72, barnes92}, commonly found in
both the field and in clusters (although more efficient in the
field). Energetic feedback from active galactic nuclei/supernovae
(AGN/SN) also contributes to star formation quenching
\citep{springel05, hopkins07}, but how much and in which ways is still
a matter of debate. These processes should be more or less efficient
at different mass scales \citep{Kaviraj2007}.
While AGN feedback is intriguing because it can justify several relationships between the properties of the central supermassive
blackholes and their host properties, finding evidence of star formation quenching by AGN activity is challenging and it is not clear whether it is an efficient and ubiquitous process. 

The first complexity in investigating this topic is related to the collection of a reliable sample of sources hosting AGNs. They show a variety of properties, and the identification rate -- thus the role of AGN feedback -- depends on the selection criteria. 
Deep X-ray surveys have provided the most effective method of
identifying reliable and fairly complete samples of active galactic
nuclei out to high redshifts \citep[\eg, see][for a review]{Brandt2005}, but they suffer from incompleteness against the most obscured sources when Compton-thick mechanisms are at work \citep[][]{DellaCeca2008,Comastri2011,Brightman2014}. Unfortunately Compton-thick AGNs represent a sizable fraction of the full AGN population, of the order of 35\% of the entire AGN population at all redshifts \citep{Akylas2009, Vignali2010, Alexander2011, Buchner2015}.
Another promising technique to obtain a reliable census of AGNs is that based on optical emission lines. Diagnostics based on emission line ratios evaluating the excitation mechanisms of the emitting gas have been used in low-$z$ optical surveys \citep{Baldwin1981}. 
At intermediate redshifts the diagnostics based on the Mass-Excitation
and the colour-Excitation efficiently separate AGNs, star forming and composite galaxies \citep{Juneau2011,Juneau2014,Yan2011}. Other approaches using a combination of different spectral features have been used at higher-$z$ \citep[\eg, the \oii\ and \hbeta\ vs \oiii/\hbeta\ by][]{Lamareille2010}. 
However, they all mandate a full coverage of the spectral window from \oii\ up to \oiii, thus limiting the selection to a small redshift range.
 
The size and the nature of the galaxy samples are other complexities
arising in the study of star formation quenching by AGN
activity. Investigations of AGN feedback are often based on transition galaxies. 
Among them, the most generally used are galaxies in the green
valley, which are interpreted to be in a transitioning stage between the blue
cloud and the red sequence \citep{Martin2007}. 
Although independent both on the type of quenching process and on galaxy mass \citep{Trayford2016}, catalogues of green galaxies selected with photometric criteria are normally scanty. 
The dearth of points in this region of the colour-colour diagram is explained by the short time required by a galaxy to cross the green valley (up to 1$-$2\,Gyr).

If photometric criteria are not producing sizable catalogues of
transiting galaxies, those selected spectroscopically suffer from
similar complexities. This class of galaxies - dubbed as
post-starburst galaxies - exhibit strong Balmer absorption lines and
no, or very faint emission lines \citep{Dressler1983}. This
combination of peculiar spectral features are indicative of a major
burst of star formation activity that has been recently terminated. Apart from the scarcity of these sources, as pointed out before, there is the need to have sufficient signal-to-noise in the spectra. It has also been noticed an incompleteness towards quenching galaxies with nuclear activity when spectroscopic selections are used. 
As the AGN heats the cold gas photoionized by hot massive stars in star-forming regions, the same emission lines become ionized by both (SF and AGN) sources \citep[\eg][]{kau03a, Davies2014}. The limit imposed on the nebular emission lines powered by both SF and AGN sources that follows the definition of a post-starburst galaxy, precludes the inclusion of these galaxies in the search. 
This underestimation is more severe at intermediate-redshifts than at low ones, because the [OII] line being routinely used at $z>0.5$ (instead of the H$\alpha$ line) particularly suffers from this problem \citep{Yan2006}. The same authors note that a less conservative cut on the [OII] line does not mitigate the problem as it produces a high contamination from other categories of galaxies, \ie, star forming objects. 

To overcome the above-mentioned shortcomings we use a photometric
criterion to select transition galaxies from the largest spectroscopic
survey currently available at intermediate redshifts, the VIMOS Public
Extragalactic Redshift Survey,
\citep[VIPERS;][]{Guzzo2014,Garilli2014,Scodeggio2017}. The
spectroscopic data are used to trace AGN activity and investigate
whether it plays a role in regulating the star formation activity in
the host galaxy by analyzing star-formation tracers like the \oii\
luminosity, as well as stellar mass and age.

The existence of the \nevl\ emission line in a galactic spectrum implies the presence of hard radiation with photon energies above 96.6~eV, that is in the extreme-ultraviolet and soft X-ray range. 
Other works have used the high-ionization potential of the \nevl\ emission line to establish the presence of gas photoionized by an AGN \citep{Feltre2016,Mignoli2013}. 
The selection based on the detection of the \nev\ line yields a
highly reliable AGN sample~\citep{Gilli2010,Mignoli2013}. 
In VIPERS it offers a unique opportunity to obtain a large sample of galaxies hosting an AGN activity at optical wavelengths over the nearly entire redshift range covered by this survey. 

In this work, we present a sample of 529 galaxies at redshift $0.62\le
z \le 1.2$ with the high-ionization narrow emission line \nevl\ as a
tracer of AGN activity in the VIPERS spectra. These candidates are selected to examine the (negative) AGN feedback in (quenching) the star formation activity in the host galaxy. 
Out of the scope of this paper is a detailed discussion on the methodologies
  of selecting AGNs. In addition, we do not include in the analysis
  those AGNs where the star formation is not the predominant source of
  emission (type\,1 AGNs), because their optical spectra being dominated
  by AGN emission, makes studying the stellar populations of the host too uncertain. 

The present work is organized as follows: the data and sample selection are presented in Sect.\,2; results are described in Sect.\,3, and a discussion and conclusions are given in Sect.s 4 and 5. Throughout this work, we assume a standard cosmological model with $\Omega_M = 0.3$, $\Omega_\Lambda = 0.7$, and $H_0 = 70 \, \mathrm{km} \,\mathrm{s}^{-1} \, \mathrm{Mpc}^{-1}$. Magnitudes are given in the AB system.

\section{Selection of \nev\ emitters}
\label{sec:sample}

\subsection{General overview of the survey}

The VIPERS project is to date the largest redshift survey of galaxies
in the  redshift range $z=[0.5-1.2]$. Its target galaxies have been
selected from the Canada-France-Hawaii Telescope Legacy Survey Wide
(CFHTLS-Wide, Mellier et al. 2008) over the W1 and W4 fields (see
Guzzo et al. 2014) and probes a volume of $\sim 1.5\times 10^8\,{\rm
  Mpc}^3 {\rm H}^{-3}_{70}$ for a total of 24~deg$^2$. The survey is a
combination of a flux-limited ($i_{AB}<22.5$) sample with $(u-g)-(r-i)$ colour selections to focus on galaxies at intermediate redshifts and comprises about 100,000 redshifts.
 
The VIPERS objects are observed with the red $R\sim 210$ VIMOS LR
grism covering the spectral range 5500--9500\AA. All details of the
observations and data handling are contained in Guzzo et al. (2014),
Garilli et al. (2012, 2014), which also describe other important
aspects of this survey. In particular, we have adopted the observing
strategy proposed by Scodeggio et al. (2009) to maximize the number of objects to be observed in a single pass using minimal slit length. Thanks to this choice the target sampling rate (the ratio of observed targets over the total number of targets) is $\sim$ 45\%. In this work, we have used galaxies with high-quality spectroscopic flags (with $>95.5\%$ confidence level, see Guzzo et al. (2014)) for source selection. 
We describe the other relevant quantities used in this work, like the stellar mass, magnitudes, and line fluxes in Sect.s 2.2 and 2.3, and we refer the reader to the following papers for fully explained details \citep[][]{Guzzo2014,Garilli2014,Scodeggio2017,Moutard2016b,Moutard2016a,Davidzon2013,Davidzon2016}.

\subsection{[NeV] emitters and their parent galaxies in the NUVrK diagram}

We use the rest-frame (NUV-r) versus \rk\ diagram, or NUVrK diagram (Fig.\,1) to select our sample of green galaxies and the corresponding control samples of red and blue galaxies \citep{Ilbert2013,Davidzon2016}. 
The NUVrK diagram is an efficient alternative to the UVJ diagram \citep{Williams2009}, and better separates the star-forming galaxies and quiescent objects as described in \citet[][]{Arnouts2013,Moutard2016a,Moutard2016b}.
We define the green valley as the locus satisfying the following colour criteria:

\begin{align} 
&{\rm for}\,\, (r-K)\le0.4 \nonumber \\
&\,\,\,\,\,\,\,\,\, 3.13<({\rm NUV} - r) <3.73 \nonumber \\
&\\
& {\rm for}\,\, 0.4<(r-K)<1.35\nonumber \\
&\,\,\,\,\,\,\,\,\, 1.37 \times (r - K) + 2.6 < ({\rm NUV} - r) < 1.37 \times (r - K) + 3.18\nonumber 
\end{align}

by using absolute magnitudes and stellar masses following the methodology used in \citet{Moutard2016b}. To account for colour uncertainties we exclude a region of 0.25~mag around the green valley locus.

The galaxies in the NUVrK diagram above and below the extended green valley region, defined as in Eq.~1 and then
expanded in all directions by 0.25\,mag (grey solid lines in Fig.~1), represent the comparison samples of the red (quiescent) and blue (star-forming) galaxies, respectively.

We use the high-ionization potential of the \nevl\ emission line to reveal the presence of gas photoionization by AGN \citep{Schmidt1998,Gilli2010,Mignoli2013,Feltre2016}. 
We define the \nev\ emitters as galaxies showing a \nevl\ line satisfying the following conditions: 

1) The FWHM of the line is between 7 and 22\AA; this is equivalent to
requiring the full line width to be from one to three resolution
elements given the resolution of VIPERS spectra. These values have
been defined based on several tests carried out on the main emission lines observed in the VIPERS spectra, a random subsample of which were visually inspected.
  
{2)} The line flux is detected at $\ge 2\sigma$. Here the flux is computed as the integral of the Gaussian best fit to the emission line, and the error on the flux takes into account the error on the continuum, the Poissonian error on line counts, and the Gaussian fit residuals. 

When the line is not detected, we have computed the upper limit to the
flux as 3 times the rms noise of the continuum adjacent to the \nevl\ line. Considering both detections and upper limits, our sample is 
representative of \nevl\ emitters having line luminosities larger than $2.1\times 10^{40}$ \lum\ with median equivalent width for the \nev\ line of about 7\AA\ over the redshift interval z=[0.62-1.20].

\subsection{Selection of representative catalogues}

As all other flux-limited surveys, VIPERS suffers from the classical selection biases induced by the flux limit selection. The common methodology to overcome this kind of selection bias is to impose a luminosity or mass cut, deriving luminosity or mass complete samples. 
Within VIPERS, the stellar mass completeness limit is \logm$ = 10.89$ over the redshift interval $z=[0.62-0.90]$ and \logm$ = 11.20$ above redshift $z=0.9$ (Cucciati et al. 2017). 
Applying the mass completeness cut at $z=[0.62-0.90]$, our sample would be composed of 1435, 719, and 528 galaxies in the red, green, and blue regions, respectively. Unfortunately with such a drastic cut, the number of \nevl\ emitters becomes very low, leaving only 32, 17, and 41 galaxies in the three colour classes.
Reducing the redshift interval down to $z=[0.62-0.80]$, and thus lowering the stellar mass completeness limit to \logm$ = 10.66$, 
the number of \nevl\ emitters remains at a similar low level. 

In order to have a robust, and reasonably large catalogue of \nevl\
emitting galaxies, that is not hampered by selection biases due to the survey flux limit selection criterion and still suitable for properly comparing the properties of these \nev\ emitting galaxies with those of the parent samples of non-\nev\ emitters, we adopt the matching technique on both redshift and stellar mass. 
With this methodology, we can obtain subsamples of galaxies with equivalent distributions of stellar mass and redshift, while spanning a wider interval of galaxy properties compared to the adoption of a pure stellar mass completeness criterion. 
The procedure firstly divides the samples of red, green, and blue
galaxies defined in Eq.\,1. into cells defined in the stellar mass
vs. redshift plane. As we are interested in the green galaxies, we
take them as reference sample. For each reference cell 0.25~dex wide
in stellar mass and 0.20 in redshift, we extract subsamples of red and
blue galaxies that match the stellar mass and redshift distributions
of the green galaxies in the corresponding cells. Thanks to the
dimensions of the VIPERS survey, we can afford to double the size of
the red and blue galaxy subsamples in each cell with respect to that of the green galaxies. 
The total number of objects in the matched stellar mass$-$redshift
subsamples is 2636 green galaxies and 5272 galaxies in each of the red and blue catalogues, covering the redshift range z=[0.62-1.20].

We repeat the matching algorithm so to have 20 subsamples of galaxies with equivalent stellar mass and redshift distributions for the comparison populations of the red and blue galaxies.
We use these 20 subsamples to evaluate the robustness of the results
for the red and blue samples. In all of the following figures, we plot the median values of the properties of these subsamples, while the error bars are the median absolute deviations. For the green galaxies, which are our reference sample and for which we cannot use the multiple catalogs extraction method, the quoted error are Poissonian errors. Finally, from each matched sample we extract the subsample of \nev\ emitters, which are the subject of our investigation.
The matched samples contain $95\pm9$ ($3.6\%$) green \nev\ emitters, {$158\pm2.5$} ($3.0\%$) red ones, and $276\pm11$ ($5.2\%$) blue ones. Using the mean instead of the median, the number of blue and red \nev\ emitters does not change significanty.

The subsamples of \nev\ emitters in the matched catalogs are plotted in Fig.\,1 with symbols colour coded accordingly to their colour-selected classes.
The sample used in each visualization of this work is the one among the 20 subsamples showing properties closest to the median properties of the 20 multiple extraction subsamples.

Following \citet{Davidzon2013}, in the estimate of the percentages
presented in this work, we take into account the incompleteness
resulting both from non-targeted sources and from spectroscopic
failures. We do not need to take into account incompleteness due to
the colour cuts (Cucciati et al. 2017), as it applies only to
redshifts lower than $z=0.6$. We have also verified that the results
obtained for the distribution corrected for non-targeted and
unidentified sources does not change compared to the observed
distribution. As expected the unobserved sources do not occupy a
special position with respect to the properties investigated here. 

\begin{figure}[t!]
\begin{center}
{\includegraphics[width=9.1cm,angle=0]{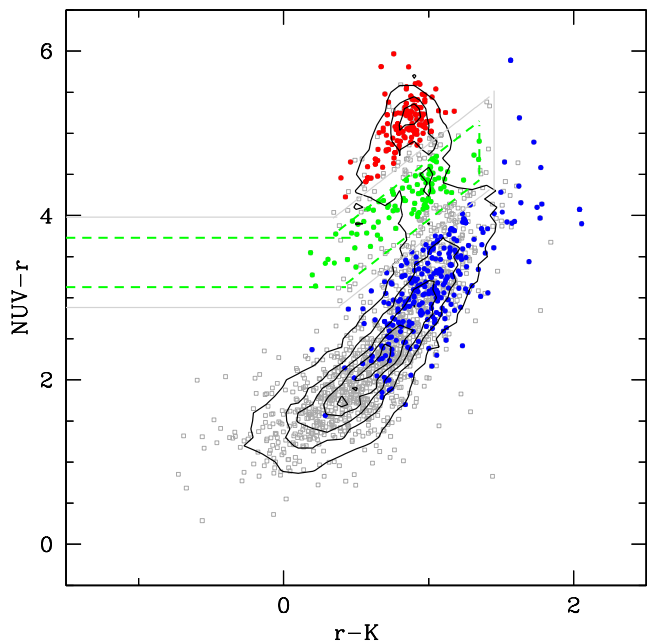}} \end{center}
\caption{Rest-frame (NUV-r) vs. \rk\ colours for VIPERS galaxies at
  $z=[0.62-1.2]$. Isophotal contours outline the loci of the whole
  VIPERS flux limited sample (contours are in steps of 10\% with the
  faintest isophote level starting at the 10\% level). The gray open
  circles represent all \nevl\ emitters in the flux-limited
  catalogue. The dashed green lines show the region of the green
  valley as defined in Eq.\,1. This is surrounded by a 0.25\,mag wide region which we have excluded to account for colour uncertainties.  The comparison samples of the red and blue galaxies occupy the regions above and below the solid gray line, respectively. The filled circles - colour coded accordingly with their classes, show the \nevl\ emitters extracted from the stellar mass$-$redshift matched subsamples.}
\label{fig:fig1} \end{figure}

\begin{figure}[t!]
\begin{center}
{\includegraphics[width=9.1cm,angle=0]{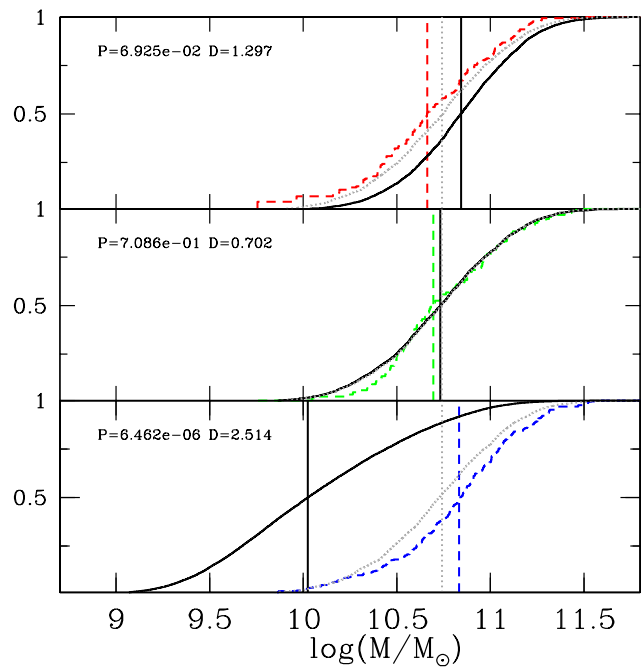}} \end{center}
\caption{Stellar mass cumulative distributions in the matched samples
  (dotted, gray lines) for the red, green, and blue galaxies from top
  to bottom. The stellar mass distributions in the flux-limited
  samples (solid black lines) and that of the \nev\ emitters
  (extracted from the matched samples and shown with dashed lines, colour-coded accordingly) are also shown together with their median values (vertical lines and same style as before). 
The Kolmogorov-Smirnov test (probability and distance quoted in each
panel) confirms as statistically significant the difference between
stellar mass distributions of \nev\ emitters and their parent matched samples in the blue cloud (but not in the green and red subsamples).
The blue \nev\ emitters are significantly more massive than the red \nev\ emitters (cfr. dashed lines in the top and bottom panel).} 
\label{fig:fig2} \end{figure}

\section{Analysis and results}
\label{sec:results}

To obtain a coherent picture of the nature of the selected \nevl\ emitters in the context of the AGN feedback scenario, we compare their properties against those of galaxies in the blue cloud, green valley, and red sequence.

\subsection{Stellar mass distribution}

Figure\,2 shows the cumulative distributions of the stellar mass for the red (top), green (middle), and blue galaxies (bottom). The black lines are the mass distributions of the full flux-limited samples, the grey dotted lines are the mass distributions of the stellar mass$-$redshift matched samples, while the coloured dashed lines are the mass distributions for the \nev\ emitters extracted from the matched samples. The vertical lines represent the median values of the stellar mass distribution of these subsamples. The stellar mass distributions of the stellar 
mass$-$redshift matched samples are equal by construction, the median value being \logm $\,=10.74 \pm 0.02$.

The mass range covered by the stellar mass$-$redshift matched samples for the
red and the blue galaxies is narrower than the mass range covered by
the full flux limited sample of red or blue galaxies. This is
intrinsic to the matching technique, which excludes from the matching
catalog objects with stellar masses not represented in all the three
colour-selected samples at equivalent redshifts. As a consequence,
also the \nev\ emitters with very low or very high mass are excluded
from the red and blue matched samples. This is particularly noticeable
looking at the \nev\ emitters of the blue cloud in Fig.\,1: a long
tail of \nev\ emitters with very blue colours that have stellar mass
values below the lower mass limit of the stellar mass$-$redshift matched
samples are excluded from the analysis (see Fig.\,2 bottom panel,
solid black vs. dotted grey curves).

The median, as well as the distribution, of the stellar mass of the galaxies in the green valley and of the green \nev\ emitters are comparable within the estimated errors,  \logm$_{green}\,=10.74 \pm 0.02$ and \logm$_{[NeV]}  =10.70 \pm0.02$ (see the grey dotted and green dashed vertical lines in Fig.\,2, respectively). 
A similar result is found for the galaxies in the red sample: both the red objects and the red \nev\ emitters have similar stellar mass distributions and the median values are consistent within the uncertainties. 
On the contrary, the stellar mass distribution of the blue \nev\ emitters is statistically different from the parent sample (\logm$_{blue, NeV} =10.88 \pm 0.02$).

{It is remarkable that the \nev\ emitters extracted from samples with equivalent stellar mass distributions (as shown in Fig.\,2 dotted lines of each panel) are significantly more massive in the blue cloud than their red counterparts at comparable redshifts (cfr. dashed lines in the top and bottom panel of Fig.\,2).}

\subsection{Stellar ages}

\begin{figure}
{\includegraphics[width=9cm,angle=0]{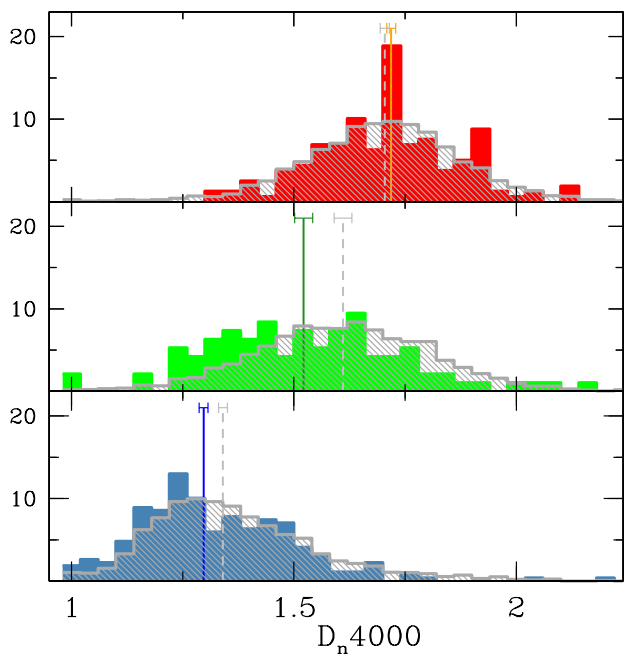}}
\caption{The \dn\ index distributions for the red, green, and blue
  \nev\ emitters plotted as filled coloured histograms (from top to
  bottom panels, respectively). The dashed, gray histograms represent
  the control samples of each population. The median \dn\ value for
  each class is overplotted (solid coloured line for the \nev\ emitters and dashed gray line for the comparison samples). The errorbars plotted at the top of the vertical lines represent the median standard deviation of each distribution. The \nev\ emitters of the green valley and those of the blue cloud have younger underlying stellar populations compared to their parent populations.}
\label{fig:fig3}\end{figure}

The strength of the 4000\AA\ break (\dn) can be used as an estimator
of the age of stellar populations \citep[\eg][]{Hamilton1985}, and
using the narrow \dn\ definition \citep{bal99} further reduces the effect of dust reddening \citep[see for details][]{kau03a}. 
Figure\,3 shows the \dn\ distribution for the \nev\ emitters (coloured filled histograms) and for the stellar mass$-$redshift matched comparison samples (grey dashed histograms) for the red, green and blue populations. Solid and dashed vertical lines indicate the medians of the different distributions. The errors computed using the resampled catalogs are overplotted. 
Being systems with absent or negligible star-formation activity, red galaxies have old underlying stellar populations, typically described by large 4000\AA\ breaks \citep[e.g.][]{kau03b,Vergani2008}. In our red sample the median value is ${\rm D_n4000_{red}}=1.70 \pm 0.01$.
On the contrary, galaxies in the blue cloud are actively forming stars
and are characterized by low values of the 4000\AA\ break, ${\rm
  D_n4000_{blue}}=1.34\pm 0.01$ \citep[see also][for these trends in
the VIPERS sample]{Haines2017}. The green galaxies, consistent with
the interpretation of being a population in transition between the blue cloud and the red sequence, show intermediate values of the 4000\AA\ break, ${\rm D_n4000_{green}}=1.61 \pm 0.02$. 
Interestingly, we find that the blue and green galaxies with nuclear activity traced by \nev\ emission show statistically significantly lower \dn\ values compared to their parent samples (${\rm D_n4000_{green, neon}}=1.52 \pm 0.02, {\rm D_n4000_{blue, neon}}=1.30 \pm 0.01$). Instead there is no statistical difference between the red \nev\ emitters and their parent galaxies of the red cloud (top panel of Fig.\,3, ${\rm D_n4000_{red,neon}}=1.72 \pm 0.01$). 
It is reasonable to assume the \dn\ spectral index to be a tracer of the stellar ages even in the presence of an obscured AGN. The AGN host colors and \dn\ values are not expected to be significantly affected by the AGN light \citep{Wang2017,Pierce2010}.
In addition, we recall here that the AGN is not the dominant mechanism in our sources, and each optical spectrum has been visually inspected in the process of redshift determination when we assign also a special flag for Type\,1 AGN (not included as explained in Sect.\,1). Under this assumption blue and green \nev\ emitters have a younger stellar population compared to galaxies of the blue cloud and green valley. 

{Previously in Sect.\,3.1, we found that the blue \nev\ emitters to be
  more massive than their parent galaxies while the red \nev\ emitters
  are slightly less massive (but within the errors). The blue and red \nev\ emitters thus have very different stellar mass distributions.
Given the correlation between the \dn\ index and stellar
mass for galaxies within the blue cloud \citep{Haines2017} or on the
red sequence \citep{Siudek2017}, it is important to verify whether the reported differences in the \dn\ values between the \nev\ emitters and their parent galaxies (in the green valley and in the blue cloud) remain statistically significant assuming a similar stellar mass distribution.
Therefore we build new catalogues using a different approach so to have \nev\ emitters and their parent galaxies of the three colour-selected classes with equivalent distributions in both stellar mass and redshift.
However, being the blue \nev\ emitters more massive than their parent sample, they would have a lower \dn\ when lowering the stellar mass distribution; this would enhance the observed behaviour. Nevertheless, for the sake of consistency we select all \nev\ emitters from the flux-limited sample of galaxies in the red sequence, green valley, and blue cloud. From these, we create redshift and stellar mass matched catalogues of \nev\ emitters as well as of non-\nev\ emitters, following the same procedure outlined in Sect.\,2.3. In other words, instead of using all green galaxies as reference, building stellar mass$-$redshift matching samples, and then extract the \nev\ emitters from these, we use all \nev\ emitters and parent samples, and then build the matched samples. We find that our conclusions are unaffected by the possible caveat induced by the different stellar mass distribution of \nev\ emitters. Using these new catalogs, the 4000\AA\ Balmer break median value is lower for blue \nev\ emitters (${\rm D_n4000_{blue, neon}}=1.22\pm0.01$, compared with 1.30 we obtained before) as well as for blue galaxies (${\rm D_n4000_{blue}}1.29\pm0.01$ compared with previous value of 1.34), but the difference between them remains significant at more than 3$\sigma$ level. Equally significant remains the difference reported for the green objects, and unchanged the \dn\ values for the red ones. }
We can safely conclude that the \nev\ emitters of the green valley and those of the blue cloud have a younger underlying stellar population compared to their parent population when using as tracer the 4000\AA\ Balmer break.

\subsection{Fractional abundances of [NeV] emitters}

\begin{figure}
\begin{center}
{\includegraphics[width=9cm,angle=0]{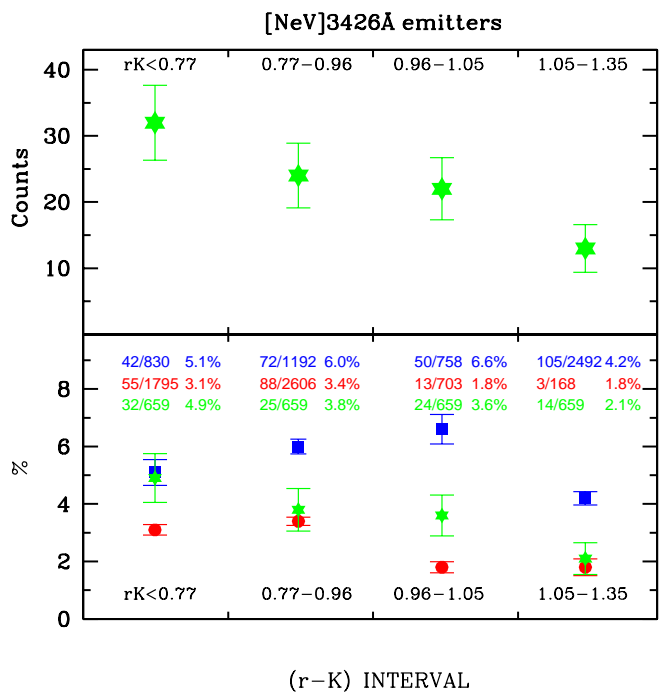}}
\end{center}
\caption{Top panel: The absolute numbers of \nev\ emitters within the
  green valley, computed in four equally-populated intervals of the
  \rk\ colour. Error bars represent the Poissonian uncertainty.
Bottom panel: Fractions of \nev\ emitters among the red sequence
(red circles), green valley (green stars) and blue cloud (blue
squares) galaxy populations. 
Error bars on the blue and red points are computed using the resampling technique. 
The fraction of \nev\ emitters in the green valley is preferentially higher
in galaxies with bluer (r-K) colours.}
\label{fig:fig4}
\end{figure}

\citet{Schawinski2014}, and references therein, claim that there are
two populations of galaxies in the green valley, with different
global properties and pathways to the red sequence. Using the VIPERS
Multi-Lambda Survey, \citet{Moutard2016b} identify as fast-quenching
objects those galaxies in the NUVrK diagram with very blue colours \rk\, $<0.76$. 
These fast quenching objects are mostly characterized by young stellar populations and feed the low-mass part of the red sequence.
In summary, the majority of the quiescent population has as progenitors evolved massive star forming galaxies with \rk $> 0.76$, while a small fraction is the result of a fast quenching process, involving AGN activity, occurring for blue and low mass objects.

To investigate the impact of AGNs in quenching the star formation
activity in galaxies, Fig.\,\ref{fig:fig4} shows the relative
frequency of \nev\ emitters as a function of their \rk\ colour. In the
top panel we show the number of green \nev\ emitters in four \rk\
colour bins (indicated in the figure). The four bins have been chosen
so that they contain equal numbers of green galaxies from the matched samples. In the bottom panel, we show instead the fraction of \nev\ emitters in the matched samples of the three galaxy populations. 
The total number of \nev\ emitters in each colour bin is annotated along with the total number of objects, and their fractions for the blue, red, and green galaxies, respectively.

\noindent The top panel of Fig.\,4 shows that {the absolute number of
  the \nevl\ emitters in the green valley becomes progressively larger
  towards bluer \rk\ colours, increasing by about a factor of 2.3 from
  the reddest bin to the bluest bin}. 
In the bottom panel the fraction of green \nev\ emitters compared to the parent samples goes from 2.1\%$\pm 0.5$\% in the reddest \rk\ bin up to 4.9\%$\pm 0.4$\% in the bluest region of the green valley. 
The same increase in the fraction of \nev\ emitters in the green valley towards bluer \rk\ colours is observed at lower {4000} Balmer break values, ranging from 2.1\% for \dn$>1.74$ up to 6.2\% in the region of the green valley with \dn$<1.48$.
No clear trend is observed in the other two classes of the blue and red galaxies which have on average a fraction of 5.5\% $\pm 0.2$\% and 2.5\% $\pm 0.1$\% \nev\ emitters, respectively.
 
\begin{figure}
\begin{center}
\resizebox{\hsize}{!}{\includegraphics[width=0.8cm,angle=0]{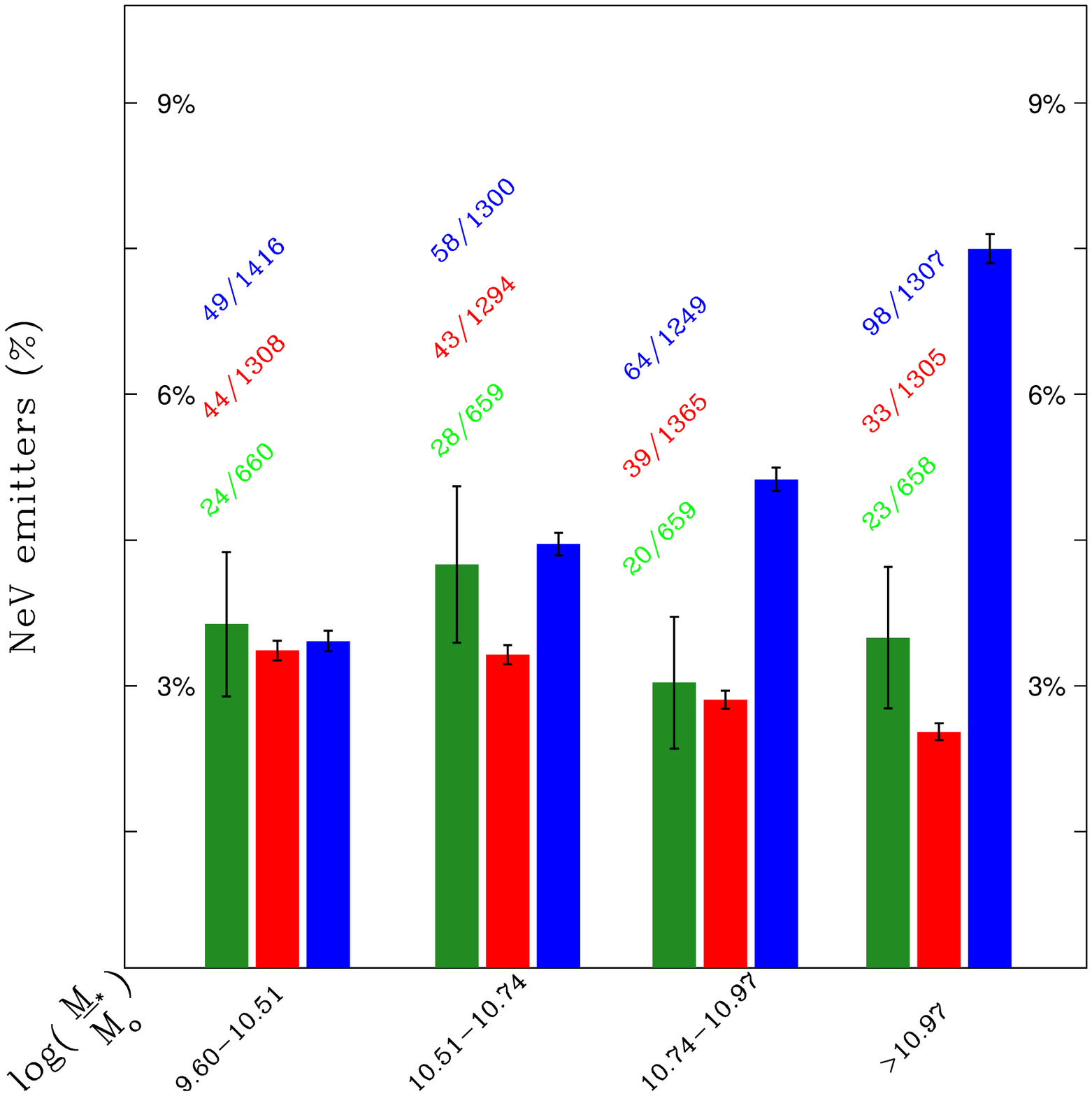}}
\vskip -3.7cm {\includegraphics[width=9cm,angle=0]{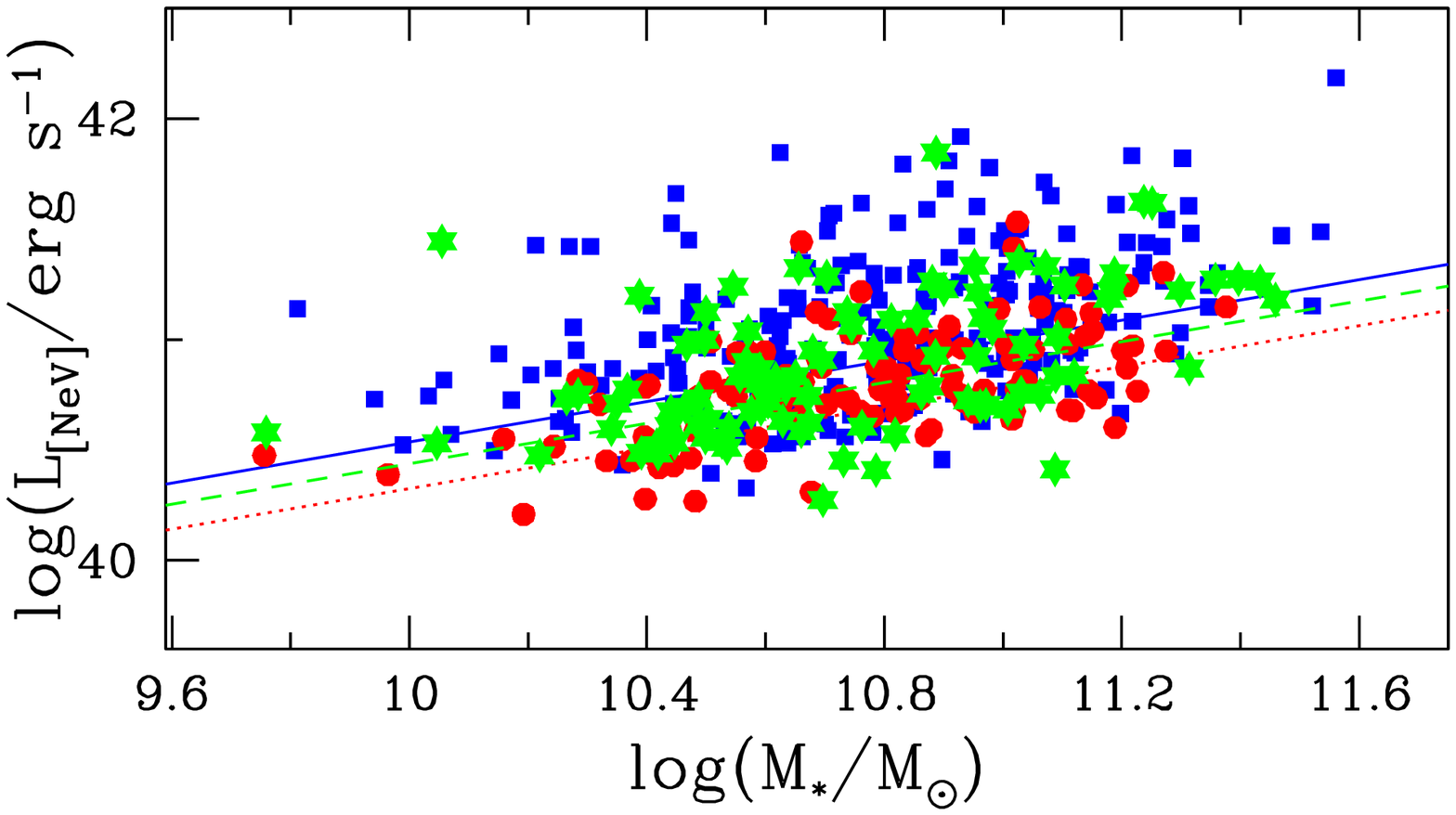}}
\end{center}
\vskip 0.5cm
\caption{(Top) The percentage of \nev\ emitters as a function of
  stellar mass for the green, red and blue galaxies. The fraction of
  the \nev\ emitters increases with mass in the blue cloud and decreases in the red sequence, but no statistical trend is observed in the green valley. (Bottom) Luminosities in the \nev\ line as a function of stellar mass. Symbols as in Figure 4.
Solid, dashed, and dotted lines show the least squares fits to the blue, green, and red \nev\ emitters, respectively. An increasing \nevl\ line luminosity is observed at larger stellar masses.}
\label{fig:fig5} \end{figure}

The increasing fraction of galaxies hosting AGNs at larger stellar mass is discussed at length in the literature \citep[][]{kau03a}. Similar trends with the quenching efficiency and stellar mass have been observed for \logm $> 10$ galaxies \citep{Kaviraj2007}. 
Figure\,\ref{fig:fig5} shows the percentage of the \nev\ emitters as a
function of stellar mass for the green, red, and blue galaxies. The
stellar mass intervals indicated in the figure represent quartiles of the stellar mass distribution in the green valley. The total number of the \nev\ emitters in each stellar mass bin and class is annotated above the corresponding histogram, along with the total number of objects in that bin. 
\noindent {The relative abundances of \nev\ emitters with respect to their parent population in the blue cloud and red sequence show opposite trends with stellar mass}. The \nev\ fraction in the blue population increases progressively with stellar mass, ranging from 3.5\%$\pm 0.1\%$ in the low-stellar mass tail up to 7.5\%$\pm 0.2\%$ in the most massive bin. 
On the contrary, the fraction of \nev\ emitters in the red sequence is
$\sim$ 3.4\%$\pm 0.1\%$ in the lowest stellar mass bin and diminishes in the high-mass bin (down to 2.5\%$\pm 0.1\%$). The green \nev\ emitters are at the level of 3.6\%$\pm 0.8\%$ showing no statistical dependence on the stellar mass. 

The bottom panel of Fig.\,\ref{fig:fig5} relates the luminosity of the \nevl\ line to stellar mass. The lines show the least squares fit to the data (solid, dashed, and dotted lines for the blue, green, and red \nev\ emitters, respectively). The blue squares represent \nev\ emitters from the blue cloud; the green diamonds those of the green valley, and the red circles those in the red sequence.
{{We observe equal increases of the \nevl\ line luminosity with increasing stellar mass for the three colour-selected classes of \nev\ emitters}}. 
The slope of the relation is $\beta=0.46\pm$0.02 in all three cases, while the intercepts decrease with redder colours, $\alpha_{\rm blue} = 35.99\pm0.03$, $\alpha_{\rm green} = 35.93\pm0.04$, and $\alpha_{\rm red} = 35.66\pm0.03$. 
{This similar dependence among the three classes of \nev\ emitters supports the intrinsic nature of the (opposite) trends in the abundance of blue or red \nev\ emitters with stellar mass (Fig.\,5, top panel). If it is simply the bulge mass defining whether the galaxy is an \nev\ emitter, we would see more \nev\ emitters in the red sequence than in the blue galaxies. Also, the bulge mass will be increasing with stellar mass within the red sequence, while the fraction of \nev\ emitters declines.

There are other findings supporting the physical nature of this behaviour: 

\noindent {1.} As found in Gilli et al. (2010), the \nev\ luminosity
correlates with the X-ray luminosity that itself correlates with the bolometric luminosity, even if a spread of $\sim1$\,dex in luminosities is observed at $\pm1\sigma$. Thus we can reasonably assume that the \nev\ luminosity is a good indicator of the AGN bolometric luminosity.
If we also assume that the Eddington ratio distribution does not
change as a function of AGN luminosity, we should detect a larger fraction of \nev\ emitters in bulge-dominated systems. 
In Fig.\,5 (top panel) we observe in blue \nev\ emitters the same trend observed in the local Universe by \citet{Bluck2014} where a larger bulge fraction has been detected in larger stellar mass galaxies over a large range of stellar masses. \\
\noindent { 2.} Furthermore, \citet[and references therein]{Bruce2016} found that lower mass \logm$<10.60$ AGN hosts have a higher mean bulge fraction than the control sample.
This study has been conducted on a sample of moderate luminosity X-ray selected AGN host galaxies at $z=[0.5-3.0]$, a population rather similar to our red population. The decrease of \nev\ emitters over the red galaxies with the stellar mass may reflect this property.}

\begin{figure}\begin{center}
{\includegraphics[width=9cm,angle=0]{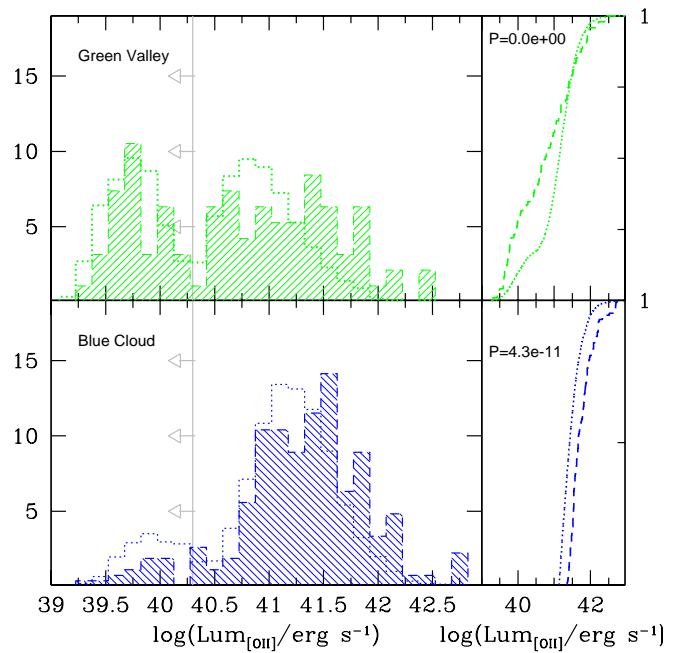}}
\end{center}
\caption{Distribution of the \nev\ (top) and \oii\ (bottom)
  luminosities (in erg/s) of the \nev\ emitters and their parent
  populations plotted in logarithmic scale. Symbols and numbers are as
  in Figs\,4 and 5, respectively. The $t$-student test and
  Kolmogorov-Smirnov test consider as extremely statistically
  significant, the difference between the luminosity (in the \oii\ line) distributions of \nev\ emitters (dashed line) and their parent samples (dotted lines). 
The vertical lines and arrows represent the luminosities where the
3$\sigma$ upper limits enter in the computation. The difference
between \nev\ emitters and their parent galaxies remains statistical
significant even if only detections are considered.}
\label{fig:fig6} \end{figure}

\subsection{[OII] Luminosity}

In Fig.\,6 we present the total (left) and cumulative (right) distributions of the \oii\ line luminosity for the \nev\ emitters (filled, dashed line) and for their parent samples (dotted line) in the green valley (top panel) and in the blue cloud (bottom panel). Both detected and upper limits to the \oii\ line luminosity are considered in this comparison. 
The vertical line in the left panels indicates at which luminosity we start dealing with 3$\sigma$ upper limits instead of detections. The fraction of upper limits in the various samples are as follows: 8.9\% blue \nev\ emitters and 16.5\% their parent galaxies; 33.7\% green \nev\ emitters and 44.2\% their parent galaxies; 57.7\% red \nev\ emitters and 62\% their parent galaxies. 
As nearly two third of the red samples are composed by non detections for the \oii\ line, from now on we perform the analysis for the blue and green galaxies only.

For both samples, the \nev\ emitters show higher \oii\ luminosities than their corresponding parent sample. Such differences are statistically significant both applying the Student's $t-$test (which gives a confidence level of more than 99.9\% for the two distributions to be different), and  the Kolmogorov-Smirnov test (which shows a nearly null probability for the two samples to be drawn from the same parent population). Such differences remain statistically significant if only detections are considered.
 {We can safely state that the \nev\ emitters show higher \oii\ luminosities than their parent galaxies in both the blue cloud and the green valley. This difference can be attributed to the AGN contribution to the \oii\ line}.

\subsection{Spectral properties}

Using stacked spectra we can investigate whether the \nevl\ emitters show  distinct features that are peculiar for this population of objects. In other words, our goal is to explore whether the presence of the \nev\ emission line, a proxy of nuclear activity, has an influence on the spectral properties of the host galaxy.

To address this question, we compare the stacked spectra of \nev\ emitters in the green valley and in the blue cloud. They have been produced by shifting each spectrum to the rest frame, and normalizing the rest-frame spectrum in the wavelength range $\lambda=3500-3700$\AA\ before median combining. Spectra with technical problems around the regions of the \nev\ and \oii\ have been removed from the stacking analysis. 
We compare the spectral characteristics of these \nev\ emitters with
spectra collected from the stellar mass$-$redshift matched catalogues
of the green and blue galaxies independent of the presence, or not, of
the \nevl\ line. Those spectra with clean regions around 3426\AA\ and 
3727\AA\ are combined following the same methodology as described above.

We observe weaker \oii\ emission lines produced by the star formation process in the composite spectrum of 70 \nev\ emitters from the green valley compared to that of 220 \nev\ emitters of the blue cloud (Fig.\,7).
The EW\oii\ is  {1.5-2} times larger in the \nev\ emitters (in the
blue and green catalogues) when compared to the stacked
spectra from their respective parent samples (1802 green galaxies and
4105 blue galaxies), supporting the results reported in the previous
section: a contribution to the \oii\ emission comes from the AGN activity, and possibly speculating an enhanced efficiency in forming new stars too. 
{The \nev\ emission line is barely detected in the stacked spectra
  from either the green valley or the blue cloud parent samples, making this line - on average - a very faint spectral feature to reveal in galaxies.}

\begin{figure}
{\includegraphics[width=9.5cm,angle=0]{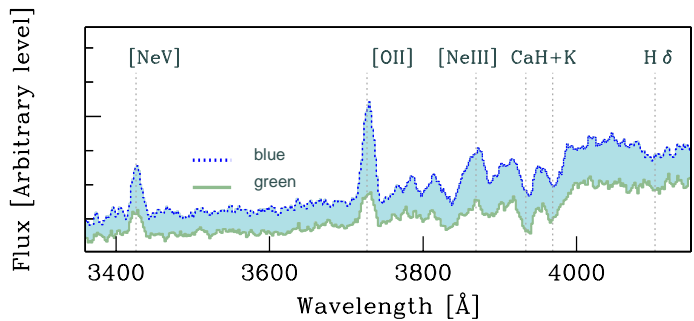}}
\vskip -4.8cm
\caption{Stacked rest-frame spectra of \nev\ emitters in the green
  valley  and blue cloud. We apply an offset to the spectra for
  clarity. The parent samples constituted by all galaxies selected
  independent of the presence of \nev\ in either the green valley or the blue cloud are not plotted.}
\label{fig:fig7} \end{figure}

\begin{figure}\begin{center}
{\includegraphics[width=9cm,angle=0]{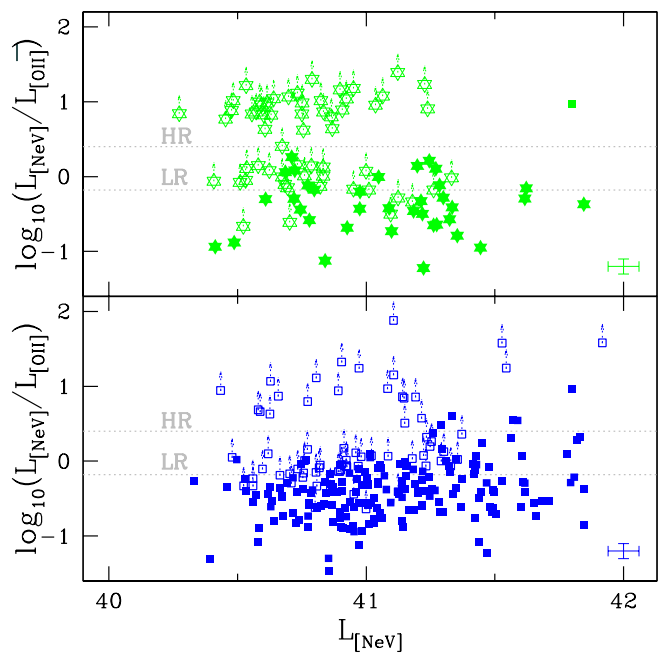}}
\end{center}
\caption{The \nev-over-\oii\ emission line ratios as a function of
  \nev\ luminosity for \nev\ emitters in the green valley (top) and blue
  cloud (bottom). Solid symbols are \oii\ detected sources. Open
  symbols and arrows represent 3$\sigma$ upper limits for \oii\ non-detections. The typical error bar associated with these quantities is plotted in the bottom right corner of each panel. The horizontal lines show the separation of each class into three equally populated regions. The combined spectra of galaxies with extreme ratios (HR and LR) are plotted in Fig.\,9.}\label{fig:fig8} \end{figure}

After a visual inspection of each spectrum used in the stack, we
identify two classes of galaxies: those galaxies with the \nev\ line
only (\nev\ pure emitters), and those with both \nev\ \& \oii\ lines (combined emitters).
\noindent Thus we separate the galaxies using a quantitative criterion based on the \nev-over-\oii\ luminosity ratio. We separate the \nev\ emitters of the green valley into three equally populated intervals based on the \nev/\oii\ luminosity ratio  (0.1-0.7), (0.8-2.5), and (2.6-24), including \oii\ lower limits. We define them as low ratio (LR), intermediate, and high ratio (HR) emitters. The same separation has been adopted to separate the \nev\ emitters in the blue cloud. We end up with 24 (HR) and 22 (LR) green \nev\ emitters, and 28 (HR) and 129 (LR) blue \nev\ emitters.\\
Figure\,8 shows the luminosity of the \nev\ line and that of its ratio over the \oii\ line luminosity for the green and blue \nev\ emitters in logarithmic scale, in the top and bottom panels, respectively. The typical error bar associated to these quantities is plotted in the bottom right corner of each panel.

Figure\,9 shows the stacked spectra of the two extreme populations with LR (panel a) and HR
\nev\ line emtters (panel b) from the green valley (green solid line) and the blue cloud (blue dotted line). The features we observe can be summarized as follows:
\begin{itemize}
\item{The stacked spectra of green and blue \nev\ emitters with low \oii/\nev\ ratios (LR) show similar spectral features, typical of star forming galaxies and AGN, with consistent line ratios (Fig.\,9a).} \\

\item{The stacked spectra of the HR blue and green \nev\ emitters show instead some clear differences (Fig.\,9b). The former shows a significant \hdelta\ absorption line (\ie, EW(\hdelta)\,$\sim$\,7.2\AA) which is absent in the HR green \nev\ emitters, and the \caii~H$\lambda3968+$\hepsilon\ is deeper than the \caii~K$\lambda3934$\AA\ absorption line (\ie, with a ratio of 0.79). The HR green \nev\ emitters exhibit properties typically associated with an old stellar population.}

\end{itemize}

To be noted that the \hdelta\  measurements should be taken as an
upper limit, due to the emission component of Balmer lines not being corrected in this study.

{To investigate whether the simultaneous presence of strong Balmer absorption lines and very faint [OII] lines in the stacked spectrum of HR blue emitters is simply due to the fact that we have imposed a low flux level for the [OII] line for these objects (\ie, \oii\ $< \frac{1}{2.6} \times$ \nev), or is instead connected with (or induced by) the presence of the [NeV] line, we have built a set of composite spectra of galaxies drawn from the green valley and the blue cloud independent of the [NeV] line detection, but satisfying the condition that their [OII] luminosity distribution is statistically consistent with that of the HR [NeV] emitters.}

Thanks to the large statistics of VIPERS, we can afford to build a
parent sample with these characteristics that is ten times larger than the  \nev\ counterparts.
These parent galaxies (from here on, the \oii\ parent population) have
equivalent distributions of the stellar mass, redshift, and \oii\
luminosity of the HR emitters. These matching catalogues are built
separately for the green valley and the blue cloud populations based on their own properties. We produce stacked spectra for this parent population following the same prescription described above.

In panel c) of Fig.\,9 the blue dotted line shows the stacked spectrum of the HR blue \nev\ emitters (same as panel b) while the gray solid line show the stacked spectrum of the blue \oii\ parent population. We can see that the \oii\ parent population does not show the inverted CaII\,H\&K ratio, while the \hdelta\ absorption line is less important (EW(\hdelta)$=2.9\pm0.4$\AA). 
This difference suggests that the peculiar spectral properties seen in the blue HR \nev\ stacked spectrum are associated with the presence of the \nev\ line and thus with the AGN activity. 

\begin{figure*}[th!]
\begin{center}
{\includegraphics[width=17cm,angle=0]{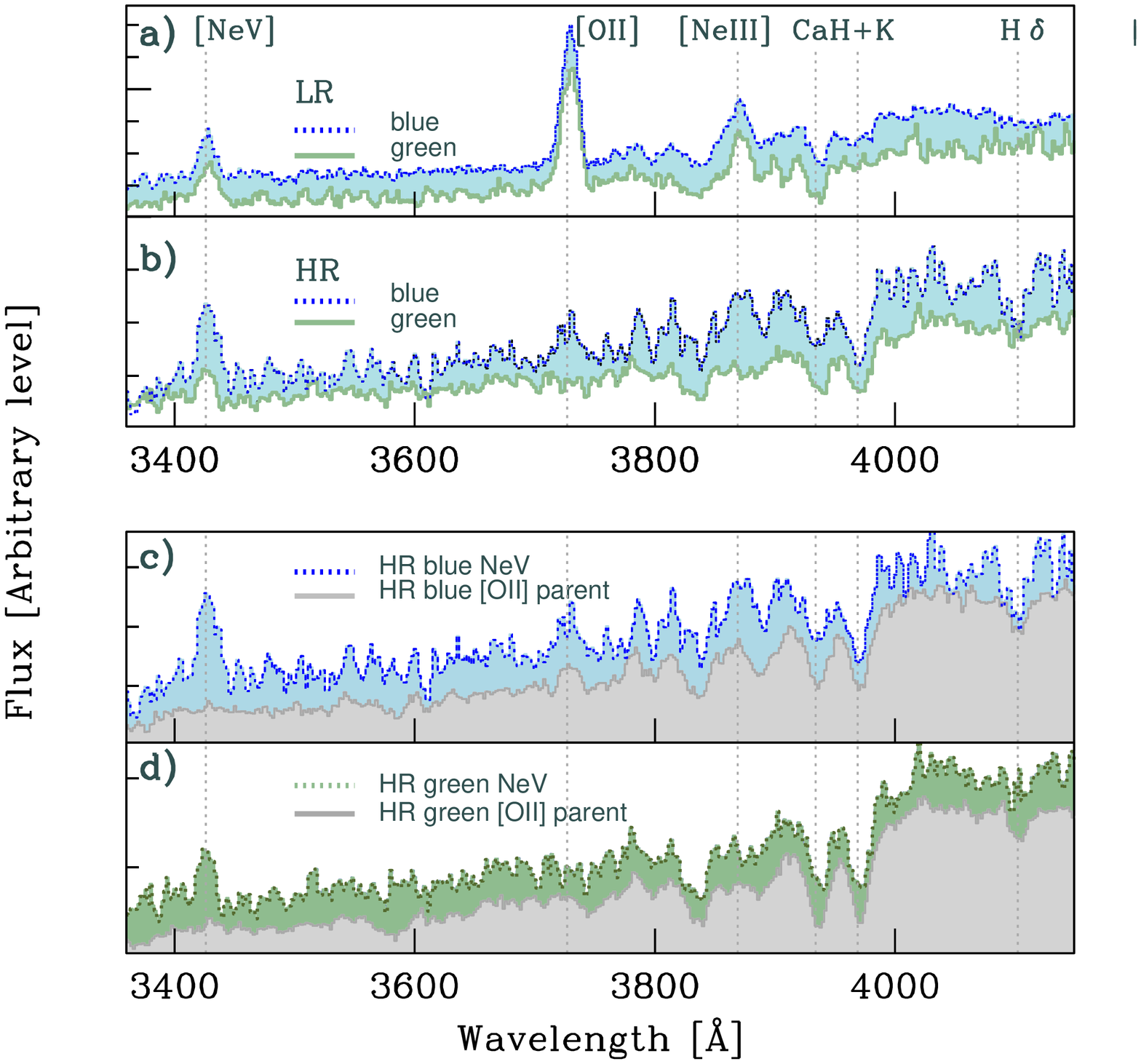}}
\end{center}
\caption{Combined spectra of \nev\ emitters with LR ({\bf a}) and HR ({\bf b}) from the green valley (green solid line) and the blue cloud (blue dotted line). 
{\bf (c)} Stacked spectrum of the HR blue \nev\ emitters (blue dotted line) overlapped to the combined spectrum of the blue galaxies (gray solid line) with equivalent distributions of stellar mass, redshift, and \oii\ luminosity. Strong Balmer absorption lines
(\hdelta, \hepsilon\ blended in \caii~H$\lambda3968$\AA) are features observed in HR blue emitters only.
{\bf (d)} Same as in panel (c) for the green galaxies and HR green \nev\ emitters.
An offset is applied to the spectra for clarity.} 
\label{fig:fig9} 
\end{figure*}

\begin{table*}
\begin{center}
\caption{\nev\ emitters $versus$ other AGN selection techniques}\label{agn_selection}
\begin{tabular}{lrr rrrr}
\hline\hline
 Sample       & N    & N in W1& N X-ray AGN (\%) &     N MIR-AGN (\%) & N MEx AGN (\%)   &  N BBPT AGN (\%) \\
\hline
 Blue \nev\   &  269 &    200 & 11 (5.5$\pm$1.7) &  9 (4.5$\pm$1.5)   & 24 (9$\pm$2)     & 21 (8$\pm$2)     \\
 Blue parent  & 5272 &   3552 & 47 (1.3$\pm$0.2) & 33 (0.9$\pm$0.2)   &112 (2.0$\pm$0.2) & 67 (1.3$\pm$0.2) \\ 
 Green \nev\  &   95 &     62 &  1 (1.6$\pm$1.6) &  1 (1.6$\pm$1.6)   &  4 (4$\pm$2)     &  4 (4$\pm$2)     \\
 Green parent & 2636 &   1708 & 13 (0.8$\pm$0.2) &  7 (0.4$\pm$0.2)   & 12 (0.5$\pm$0.1) & 10 (0.4$\pm$0.1) \\
 Red \nev\    &  157 &    120 &  1 (0.8$\pm$0.8) &  0 (0.0)           &  0 (0.0)         &  0 (0.0)         \\
 Red parent   & 5272 &   3501 & 11 (0.3$\pm$0.1) &  5 (0.14$\pm$0.06) &  0 (0.0)         &  0 (0.0)         \\
\hline
\hline
\end{tabular}\\ 
\end{center}
{\small X-ray AGN: X-ray detected sources with X-ray luminosities greater
than 10$^{42}$\,\ergs\ or with \nh$\ge10^{22}\,{\rm\,cm^{-2}}$.\\ 
MIR-AGN: sources with 5$\sigma$ detections at 3.6\,\micron\ and
4.5\,\micron\ and with S(4.5\,\micron)/S(3.6\,\micron)$\geq$1.27~\citep{Stern2012}.\\
MEx AGN: sources with more than 3$\sigma$ \hbeta\ and \oiii\ equivalent widths
and EW(\oiii)/EW(\hbeta) ratios satisfying the MEx AGN selection
criterion defined in ~\citet{Juneau2011}.\\
BBPT AGN: sources with more than 3$\sigma$ \hbeta, \oii, and \oiii\ equivalent widths
and EW(\oiii)/EW(\hbeta) and EW(\oii)/EW(\hbeta) ratios satisfying the blue
BPT criterion defined in~\citet{Lamareille2010}.}\\
\end{table*}

In panel d) of Fig.\,9 we show the same comparison for the objects in
the green valley: the stacked spectrum of the HR green \nev\ emitters (green dotted line, same as panel b) is overlaid upon the combined spectrum of the green \oii\ parent population (gray solid line). The two spectra do not show any significant differences besides the presence of the \nev\ line. We can conclude that in the green valley the presence of the \nev\ line, and thus of an AGN, does not leave an imprint on the host stellar population.

The deep \hdelta\ Balmer absorption feature is observed in galaxies
when early-type stars (A-stars) dominate the main-sequence
contribution near the epoch of the observation. The Ca\,H/K line ratio
is constant in stars later than about F02, while it increases
dramatically for earlier type stars as the \caii\ lines weaken and
\hepsilon\ strengthens \citep{rose84}. In addition to these spectral
properties, we also observe a weak \oii\ line in the stacked spectrum
of HR blue \nev\ emitters. This line is emitted in H{\small II}
regions around O and B stars, and it is an indicator of recent star
formation (with a lifetime of 10$^7$~yr). We have to consider, however, that in galaxies hosting an AGN this is only an upper limit to the real star formation level. 
As already pointed out, gas photoionized by the AGN can also
contribute to the \oii\ emission line. In such galaxies, the \oii\
emission line is the product of both sources of ionization. Based on
the weak measured \oii\ luminosity we can consider the star formation
at the time of the observation to be extremely low, if not stopped entirely. 
Together, these properties describe a class of galaxies in which there has been a very recent episode of star formation (with a burst dated back by 200-300\,Myr), responsible for the inverted Ca\,H/K line ratio,  which has now (almost) stopped (hence the extremely low \oii\ emission).

We thus simultaneously observe an AGN indicator (\ie, the \nev\ line)
and recently stopped star-forming activity (\ie, no \oii\ emission,
inverted \caii~H/K ratios, and deep \hdelta\ in absorption) in the
blue cloud sources selected on the basis of a high \nev /\oii\
luminosity ratio. Whether and how the two observables are linked is
not clear. We can however affirm that these sources, under certain
conditions linked to neither the stellar mass, nor the redshift, nor
the \oii\ luminosity, have a short quenching time, \ie shorter than
typical crossing time $\tau < 1$\,Gyr, as we cannot see their presence
in the green valley. 
Of course we cannot exclude the possibility that by the time these
sources enter and cross the green valley, they no longer have any
significant \nev\ line, thus they are not identified as \nev\
emitters. This would imply that we caught the objects during a short
phase across their complex path to the red sequence $-$ in fact we
observe that the \hdelta\ is relatively strong in the spectrum of the green \oii\ parents. 

\subsection{\nev\  emitters {\it versus} other AGN selection techniques}

In Sect.\,1, we have mentioned on other AGN selection techniques,
e.g.  based on the X-ray luminosity, mid-IR colours, the MEx diagram, or
optical emission line ratios.  These techniques yield AGN samples with
different levels of completeness and purity, and these depend on the depth of
the data.  In this work, we have adopted a less common technique for several
reasons: our selection based on the detection of the \nev\ line yields a
highly reliable AGN sample~\citep{Gilli2010,Mignoli2013}, and it can be applied to the full
VIPERS sample. In addition, numerous previous works have already
investigated the incidence of AGN identified by these techniques in the
green valley or in transitioning
galaxies~\citep[\eg,][]{Yan2006,Schawinski2010,Schawinski2014}. 
While they all find higher fractions of AGN in the green valley, there is no
clear evidence of negative feedback from these AGN.  Since different
selection techniques yield AGN samples that do not fully overlap and might
favor different evolutionary phases~\citep{Hickox2009}, it is interesting to
investigate and compare results obtained with different AGN samples.

We have conducted this analysis using the ancillary data available in the VIPERS fields to
quantify the fraction of \nev\ emitters that would be selected as AGN by the
most common selection techniques.  In particular, we have used the X-ray data
from the XXL survey~\citep{Pierre2016}, the {\it Spitzer} data from the {\it SWIRE}
survey~\citep{Lonsdale2004}, the {\it WISE} data from the {\it AllWISE}
data \citep{Wright2010}, as well as spectral measurements from VIPERS itself,
but requiring the detection of multiple emission lines, i.e.  \hbeta,
\oiii, and/or \oii~\citep[MEx and blue BPT
diagrams,\,][]{Juneau2011,Lamareille2010}.  Each of these criteria can
only be applied to subsets of the VIPERS sample, because the X-ray and {\it Spitzer} coverage is
limited to a fraction of the W1 field, and because the lines necessary to
apply the MEx or blue-BPT diagnostic diagrams are only available for objects up to $z\sim0.89$.
The numbers and fractions of \nev\ emitters selected by these techniques are listed in
Table~\ref{agn_selection}.  Overall, we find that only a small fraction,
i.e.  $<$10\%, of \nev\ emitters are selected by these techniques.  In all
cases, the fraction of selected AGN is higher among the \nev\ emitters than
in their parent samples: by a factor 5 in the blue and green
samples and a factor 2 in the red one.  This analysis also shows that the AGN
contamination in the parent samples is negligible, i.e.  $\leq$2\%.  Note
that this is true even if \nev\ emitters are present in the parent samples.

This comparison indicates that the AGN selected by the strength of the \nev\
line are not commonly represented in shallow X-ray, mid-IR, MEx, or BPT
selected samples, and any peculiarity found in this class might be linked to
what makes this line detectable.

It is interesting to note that 12 out of 13 X-ray detected \nev\ emitters
are obscured and their X-ray-over-\nev\ luminosity ratios are consistent
with those measured in other X-ray detected \nev\ emitter
samples \citep{Gilli2010}, thus confirming that \nev\ sources are mostly
obscured AGN.  However, we note that the numbers of \nev\ emitters
detected in the X-rays, or with hot dust signatures in their mid-IR
SEDs, are too low to draw any significant conclusion on their properties based on
these data \citep[see also][]{Gilli2010,Mignoli2013}.

In contrast to previous studies based on the classical AGN selection
techniques, we do not find an enhancement of AGN in the transition galaxies,
but a decreasing fraction from 5.2\%, to 3.6\%, and 3.0\% going from the
blue cloud, to the green valley, and red sequence galaxies. We also find a
different trend with respect to stellar mass; with the incidence of AGN
increasing towards higher masses in the blue cloud, but decreasing
with mass in the red sequence. Previous studies based on X-ray selected AGN find an increase
in the AGN fraction with stellar mass in the red and green samples and no variation among blue
galaxies \citep{Wang2017}.

\section{Discussion} 

In the past two decades observational studies at various wavebands support the idea of a stochastic fueling of AGN in galaxies, thus no evidence for AGN feedback on the star formation activity of the host galaxies would be needed. 
However various observational findings and theoretical studies provide then an alternative picture supporting a close connection between AGN activity and the host galaxies \citep[\eg,][]{Boyle1998,Silverman2008,Aird2010}, although this impact on galaxy formation and evolution is not yet well understood \citep{Best2006,Croton2006,Nesvadba2008,Smolcic2009}.

In this investigation we present a number of observational facts
pointing to differences between active and inactive galaxies
supporting the AGN feedback scenario. In this work the nuclear
activity is probed by the high-ionization \nevl\ emission line and
\nev\ emitters with their parent samples are properly selected
according to their stellar mass, redshift, and NUVrK colour
distributions. Broad-line (Type\,1) AGNs have been excluded because our selection focuses on galaxies with the stellar light not contaminated by continuum light from the AGNs.

We find that stellar ages and SFRs (expressed via the \oii\ luminosities) of the \nev\ emitters appear to be statistically different from the corresponding inactive galaxies once properly matched in redshift and stellar mass.
In particular, \nev\ emitters are hosted by galaxies with younger
stellar populations (lower \dn\ values) and higher SFRs (larger
\oii\ luminosities) on average than their parent population.
Stellar ages and SFRs are unchanged in active and inactive galaxies
when the AGNs are selected independently by a separation of the hosts
in colours, or the morphological aspect; differences start to be
observed instead when the samples are carefully selected, though the
results by \citet[][and references therein]{Azadi2017} have a lower
level of significance, consistent with the reduced statistics.
Consistent with previous results we find a larger fraction of green
galaxies hosting AGN activity at progressively bluer \rk\
colours. This property is in agreement with results reporting an increasing incidence of AGN activity in younger systems, from an X-ray selected sample in the COSMOS field up to $z\sim1$ by \cite{Silverman2009}.  
These different properties testify to an intimate correlation between the AGN activity and the mechanisms that regulate star formation in the host galaxy.

In addition to the different stellar ages and SFRs, we observe an
dependence of AGN fraction on the location of the host in the NUVrK
colour diagram, and that this trend is mass-dependent. 
We find a lower fraction of [NeV] emitters in the red sequence compared to that in the green valley and in the blue sequence. However, such a difference is mass-dependent as it is not observed for galaxies with stellar masses $<$3$\times$10$^{10}$\,\msun\ and it is enhanced at higher masses, i.e.  $>$9$\times$10$^{10}$\,\msun.  One possible explanation is that the AGN affects more significantly its host galaxy or viceversa when the host galaxy, and thus the accreting BH, are more massive.
The nuclear activity is also more common in progressively massive blue galaxies. 
The trend with stellar mass has been already observed at different redshifts for different types of AGNs \citep{KauffmannAGN2003, Best2005, Silverman2009}. Local AGNs selected
from the Sloan Digital Sky Survey and visually classified by
\citet{Schawinski2010} are found preferentially in massive late-type
galaxies with \logm\ $\sim11$,  and moderate-mass early-type systems with \logm\ $\sim 10$, in agreement with our results.
This is coherent with \citet{Schawinski2010}'s proposal that black holes have different properties when hosted in either early/red or late/blue type galaxies. 

Conventional line diagnostics can misidentify AGN in composite galaxies with strong star formation activity \citep[\eg][and references therein]{Bar2017}. Single, high-ionization line diagnostics are more sensitive to AGN ionization in the presence of high levels of star formation, \eg, using the [HeII]$\lambda$4685 emission line \citep{Bar2017}, or the \nevl\ emission line \citep[][including this work]{Gilli2010,Mignoli2013}.
\citet{Bar2017} report that [HeII]$-$selected AGNs (\ie not with the standard BTP diagram) are more commonly found in star-forming galaxies in the blue cloud and above the main sequence at the high masses where quenching is expected to play a role, see \citet{Haines2017}.
In this respect, the results by \citet{Schawinski2010} who locate the majority of local AGN candidates in the green region may be explained by the use of classical diagnostics. 
It is also true that if these AGNs are located in an "already
transiting" region, then the AGN cannot be the reason for initiating the quenching process, as their host galaxies should have already experienced a star formation suppression.
In fact the delay time between the star formation quenching and the detection of emission-lines used to select AGN candidates is expected to be larger than the time a galaxy needs to cross the green valley. In conclusion, it is not surprising finding quenching signatures in active galaxies of the blue cloud if the AGN is the reason of such a star formation quenching. This is infact what we find in this analysis.

We have assessed the relative importance of AGN on the star formation
activity using the ratio of the \nev\ over \oii\ line luminosities to
define specific classes of galaxies, and searching for spectral
signature of recent quenching in the stacked spectra of galaxies
belonging to each class. This methodology is similar to the ones adopted in the local Universe -- with the ratio of the IR fine structure lines by \citet[]{LaMassa2012}, or analogously to the $"D"$ parameter as distance at which a source lies from the locus of star forming galaxies on the BTP diagram by \cite{KauffmannAGN2003}.
Through this approach we identify a subset of \nev\ emitters with high
\nev\ to \oii\ ratios within the blue cloud, that show signs of recent
quenching having experienced a burst of star formation activity in
their recent past \citep[less than $200-300$~Myr ago,
e.g.,][]{Wild2007,Wild2010}. This is testified by spectral signatures typical of very hot, massive stars (Balmer absorption lines and an inversion of the intensities of the K and H calcium lines, due to the presence of a blend of the calcium H line with the ${\rm H}\epsilon$ Balmer line). 
Their spectra resemble those of post-starburst galaxies. However they would never be included in a spectroscopic search of post-starburst galaxies because of the limit imposed on the nebular emission lines $-$ powered in these galaxies by both SF and AGN activity \citep[see][and references therein]{Alatalo2016}.

There have been recent claims of a bimodality in the composition of
the population of the green valley. This bimodality has been attributed
to the different quenching mechanisms (and the characteristic timescales) in action. 
\citet{Moutard2016b} use a similar NUVrK diagram to identify green
valley galaxies within the VIPERS dataset, as a tracer of galaxy
evolution and the quenching process. They identify one main quenching
channel (over a moderately long timescale $0.5-2$\,Gyr) between the star-forming and quiescent sequences at
$0.2<z<1.5$, 
that is populated by massive, star-forming galaxies with typically red colours, \rk>0.76.
Interestingly, they also suggest a second path to the red sequence (with faster quenching timescales) that is followed by bluer, low-mass galaxies.
Similarly, in the local universe, \citet{Schawinski2014} find a
primary quenching channel of late-type galaxies that move to redder colours with
a slow quenching timescale according to the exhaustion of their gas
supplies, but also identify a minority of transitioning galaxies that requires a rapid transformation of morphology and colour. 
Our result that bluer galaxies have a higher probability of possessing
an AGN, thus a faster quenching time, and that a fraction of them present quenching signatures are in agreement with the general picture emerging in the conclusions of these studies.

This investigation demonstrates that there exists a link between 
AGN activity and the sustainability of star formation in the hosting galaxies at $0.62<z<1.2$.
The presence of the mechanism(s) producing the highly ionized \nevl\
line, under certain conditions, stops the formation of new
stars. These conditions may be related to the efficiency of the black
hole and/or the physical conditions of the hosting galaxies, but not 
linked to stellar mass, redshift, or current star formation (as expressed via the \oii\ luminosity). 
To have a completely quenched system, the cold gas reservoir has to be
either fully consumed or heated up entirely. In simulations this is achieved almost instantaneously via AGN feedback \citep{springel05}. Because the photoionization happens almost instantaneously, the requirement is to observe the AGN hosting galaxies that have not yet reached the green valley, precisely as we observe in the \nev\ emitters in the blue cloud. Such hypothesis of a rapid gas reservoir destruction has to be further investigated.
If we consider that our results are limited by the \nev\ observational limits and the rapid AGN variability, the impact of AGN feedback on galaxy formation and evolution may represent an important channel of fast-transiting galaxies to the red sequence.

\section{Conclusions}

Thanks to the large volume probed by the VIMOS Public Extragalactic Redshift Survey, we have been able to select the largest sample of [NeV] emitters at intermediate redshifts (z = 0.62-1.2) available so far. We have used this sample to study the properties of the host galaxies with nuclear
activity, as probed by the high-ionization potential of the \nevl\ emission line. The sample comprises 529 [NeV] emitters, matched in redshift and stellar mass, and divided into the red sequence, green valley and blue cloud according to their (NUV-r) vs. \rk\ colours. 
We have built matched control samples of galaxies using the full VIPERS survey, to which we compare the properties of the active sample.

We report on statistically different properties (stellar age, \oii\ luminosity, colour) between active and inactive galaxies, and among the active galaxies different characteristics (stellar mass, fractional number) are observed according to their NUVrK colours.
In particular, the main results are as follows.

{1) Blue [NeV] emitters are significantly more massive than the [NeV] emitters in the red sequence.}

{2) The [NeV] emitters in the green valley and in the blue cloud have younger underlying stellar populations compared to their parent populations (using the 4000\AA\ Balmer break as a tracer).}

{3) The absolute numbers of \nev\ emitters in the green valley increases by about a factor of 2.5 with decreasing \rk\ color. This is not observed for the [NeV] emitters in the blue cloud or in the red sequence.}

{4) The abundance of the [NeV] emitters with respect to their parent population in the blue cloud and red sequence show opposite trends with the stellar mass. No statistical trend with stellar mass is observed for the number of [NeV] emitters in the green valley.}

{5) The [NeV] emitters show a brighter [OII] luminosity than that of the parent galaxies in both the blue cloud and the green valley, confirming the contribution of the AGN activity to the [OII] emission. We also find a population of objects in which the [NeV] emission is dominant over the [OII] emission (HR emitters).} 

{6) The analysis of the stacked spectra of [NeV] emitters shows that while non-HR \nev\ emitters have similar characteristics in the blue cloud and in the green valley, blue HR emitters show signs of a recent burst of star formation (deep \hdelta\ absorption and inverted \caii~H/K ratio) while the absence (or extreme weakness) of \oii\ shows that no star formation is going on at the observation epoch.}

\medskip These observational evidences taken together point towards a statistical correlation between the AGN activity and the mechanism(s) that cause modifications of the star formation regulation in the host galaxy. 
We also reveal the existence of a novel class of AGN hosting galaxies where the  supermassive black hole may arguably  fast quench the star formation and accelerate the evolution from the blue cloud to the green valley. 

\section{Acknowledgments}
We acknowledge the crucial contribution of the ESO staff for the management of service observations. In particular, we are deeply grateful to M. Hilker for his constant help and support of this program. Italian participation to VIPERS has been funded by INAF through PRIN 2008, 2010, and 2014 programs. LG and BRG acknowledge support from the European Research Council through grant n.~291521. OLF acknowledges support from the European Research Council through grant n.~268107. JAP acknowledges support of the European Research Council through grant n.~67093. WJP and RT acknowledge financial support from the European Research Council through grant n.~202686. AP, KM, and JK have been supported by the National Science Centre (grants UMO-2012/07/B/ST9/04425 and UMO-2013/09/D/ST9/04030). WJP is also grateful for support from the UK Science and Technology Facilities Council through the grant ST/I001204/1. EB, FM and LM acknowledge the support from grants ASI-INAF I/023/12/0 and PRIN MIUR 2010-2011. LM also acknowledges financial support from PRIN INAF 2012. YM acknowledges support from CNRS/INSU (Institut National des Sciences de l'Univers). CM is grateful for support from specific project funding of the {\it Institut Universitaire de France}. SDLT and CM acknowledge the support of the OCEVU Labex (ANR-11-LABX-0060) and the A*MIDEX project (ANR-11-IDEX-0001-02) funded by the "Investissements d'Avenir" French government program managed by the ANR. and the Programme National Galaxies et Cosmologie (PNCG). Research conducted within the scope of the HECOLS International Associated Laboratory, supported in part by the Polish NCN grant DEC-2013/08/M/ST9/00664. Finally, DV acknowledges the help received by M. Sandri in computing coding.

\bibliographystyle{aa} 

\bibliography{mybiblio}

\end{document}